%% file: VolkmannCO2.tex
\documentclass[aps,12pt]{article}

\usepackage[utf8]{inputenc} % required for arXiv

\usepackage{placeins}
\usepackage{amsmath}
\usepackage[colorlinks]{hyperref} 
\input{my_commands}

\begin{document}

\title{Potentials, exchange, and correlation in attosecond photoemission delays}
\author{Hakon Volkmann$^{(a)}$\footnote{E-mail: hakon.volkmann@physik.hu-berlin.de}, Vinay P. Majety$^{(b)}$\footnote{E-mail: vinay.majety@iittp.ac.in}, and Armin Scrinzi$^{(c)}$\footnote{E-mail: armin.scrinzi@lmu.de}\\
$^{(a)}$Institut f{\"u}r Physik, Humboldt-Universit{\"a}t, Berlin, Germany\\
$^{(b)}$Indian Institute of Technology Tirupati, Tirupati, India\\
$^{(c)}$Ludwig Maximilian Universit{\"a}t, Munich, Germany\\
}

\maketitle

\maketitle
\begin{abstract}
We investigate attosecond time delays in the emission of photoelectrons using a hierarchy 
of models of the $CO_2$ molecule including the strong field approximation, Coulomb-scattering, short-range parts
of the molecular potential, Hartree and Hartree-Fock descriptions. In addition, we present an {\it ab initio} calculation
based on quantum-chemical structure in combination with strong-field techniques, which fully includes
multi-electron exchange and correlation. Every single of these model 
constituents is found to modify delays on the scale of 10 as or more, with exchange and correlation having the most 
pronounced effect.
\end{abstract}
\maketitle
\newcommand{\au}{\,au}
\section{Introduction}

Time delays in photoelectron emission are a highly sensitive probe of
the dynamics of electron ejection and subsequent scattering. Streaking \cite{Yakovlev2005} and 
RABITT \cite{veniard96:phase-dependence,paul01:atto} are two experimental techniques that can measure time delays 
with resolutions on the scale of 10 as (1 as=$10^{-18}s$). The high precision of these
measurements may uncover a variety of mechanisms that contribute to photoemission.
An interpretation in terms of ``tunneling times''\cite{P.Eckle2008} has created a long debate,
with recent measurements suggesting the absence of any such time delay in the tunnel process
\cite{sainadh19:attodelays}.
The whole debate has also been hampered by serious conceptual questions related to the  
definition and interpretation of any such delay time (see, for example, 
\cite{BuettikerM.andLandauer1982,torlina15:attoclock,Argenti17:delays,saalmann20:delays}). In course of that debate,
awareness has been heightened that irrespective of delays that might be associated with the 
photo absorption process, scattering delays dominate the measurements. 

Independent of any interpretation, the measured delays may provide a sensitive probe into 
bound-continuum matrix elements of complex correlated systems. That becomes
transparent for the RABITT technique, where a near infrared (IR) pulse is superposed with a high harmonic pulse in the extreme ultraviolet 
(XUV)  to produce side bands whose amplitudes are sensitive to the relative phase between the pulses. 
Already in the original proposal of the technique \cite{veniard96:phase-dependence} a system specific
``atomic phase'' of the bound-continuum transition was introduced. With progress of attosecond techniques, that
phase has been moved into the focus of interest and it is re-interpreted as a time delay within the optical cycle of the 
IR pulse. At sufficiently low intensities the process is a perturbative two-photon transition that solely 
depends on the initial bound and final scattering states of the field free system. This inherently 
stationary nature of the process precludes the interpretation as a time-ordered process of photo absorption followed by scattering, but it
provides a highly sensitive probe into bound and scattering solutions. 

It was noted that the picture may become more complex in presence of resonances \cite{Argenti17:delays} 
as the process needs to be studied on the level of quasi-degenerate perturbation theory. Even 
in absence of such complications, scattering solutions need to be obtained taking into account 
correlation. In a recent publication \cite{kamalov20:CO2} it was noted that for the description of delays 
in photoemission from $CO_2$ a single scattering channel is insufficient and channel coupling must be included.

In view of this complexity, fully correlated {\it ab initio} scattering solutions may be used in the 
computation of attosecond delays. This should ultimately lead to convergence of theoretical with experimental 
results. Such a verification of {\it ab initio} calculations or complex models against precision measurements may be
gratifying, but a more simplified description would be desirable for interpreting observations in terms of physical mechanisms 
such as the role of the initial state, 
system specific  single-particle scattering, the more universal Coulomb scattering, and exchange and correlation.  

In the present paper we explore which level of approximation may be admissible when the goal is to compute delays
with accuracies of $\sim 10\,as$. We quantify the effect of individual approximations using a hierarchy of closely 
related models for the $CO_2$ molecule. Anticipating the conclusions, we find that attosecond delays on that small scale 
crucially depend on every single of the effects listed above and any modeling short of the 
{\it ab initio} level is  unlikely to produce conclusive results with accuracies of $\sim 10\,as$.

\section{RABITT for the time-dependent Schr\"odinger equation }

We discuss delays that are measured by the RABITT technique, for which we summarize
the procedures here and refer to literature for a discussion, e.g. Ref.~\cite{Pazourek15:atto}. 
In that technique one measures photo emission spectra, where single-photon 
ionization by a pulse of odd high harmonics ejects an electron in presence of a weak field at the fundamental 
frequency. This generates side bands in between the odd harmonic peaks by adding a single photon of the fundamental
to one peak or, alternatively, subtracting a photon from the next higher peak. 
The interference of the two pathways causes a variation of the side-band intensity 
with the relative phase between harmonics and fundamental,
which in the time-domain can be interpreted as a delay in the harmonic photo-emission.

Specifically, we compute the 
photo-emission spectra in the combined fundamental near-IR and high harmonic fields with vector potential 
\begin{eqnarray}
\lefteqn{\int_{-\infty}^t d\tau \vEf(t)=:\vA(t)=}
\nonumber\\
&& \vA_f(t) \sin(\om t +\phi)+ \vA_h(t)\left[\sin((2n-1)\om t)+\sin((2n+1)\om t)\right]
\label{eq:vectorA}
\end{eqnarray}
at a fundamental wave length of $\la=800\, nm$ and its 13$^{th}$ and 15$^{th}$ harmonic ($n=7$).
Peak amplitudes $\max_t|\vA_f(t)|$ and $\max_t|\vA_h(t)|$ are chosen such that 2-photon perturbation theory is applicable.
In that limit, the exact amplitudes become unimportant and can be factored out. Similarly, neither the 
exact pulse shapes nor durations matter, if only the pulses are long enough to have a narrow
band-width that allows a clear separation of the harmonic peaks from the side band peak created by the fundamental.
In practice, we use pulses with FWHM duration of 3-4 $T$, where $T=\la/c\approx 2.6 fs$ 
is the optical period of the fundamental. The photon energy of the fundamental is
$\om=2\pi\hbar/T\approx1.55\,eV\approx 0.057 \au$, where atomic units (au) are defined through $m_e=\hbar=e^2=1$.

For initial state energy $E_c$ \wrt a given ionization channel, the side band appears at electron energy $E_c+2n\om$.
In a time-dependent description, the relevant part of the 
two-photon perturbative wave function at time $t$ depends on the phase shift $\phi$ between the fundamental 
and the harmonics as follows:
\begin{align}
\Psi(t) 
&=
\sin{\phi}\int_{-\infty}^t d\tau U_0(t-\tau)\cos{\om\tau}(i\vA_f(\tau)\cdot\vna) U_0(\tau) \Phi_{h}(\tau)
\nonumber\\
&+\cos{\phi}\int_{-\infty}^t d\tau U_0(t-\tau)\sin{\om\tau}(i\vA_f(\tau)\cdot\vna) U_0(\tau) \Phi_{h}(\tau).
\label{eq:twoPhoton}
\end{align}
Here we denote by $U_0(t)$ the time-evolution without the field and by $\Phi_h$ the wave packet generated
by the harmonic vector potential $\vA_h(t)$ in first order of perturbation theory.
Two time-dependent calculations suffice for determining the modulation of the side band amplitude. 
A maximal side band peak at $\phi=0$ may be interpreted as undelayed emission of the photo-electron, in which case
a maximum at non-zero $\phi$ corresponds to a delay of $\De T= T\times\phi/(2\pi)$.

As the physical interpretation of RABITT fundamentally relies on perturbation theory, it can be equivalently 
formulated using stationary scattering theory, where instead of emission delays one deals with 
phases in the bound-continuum dipole transition matrix elements. Such a formulation does not reduce the computational
challenge, as for obtaining correct phases one needs scattering solutions that are very accurate in the vicinity of the nuclei. 
We have chosen the time-dependent approach for the simple reason that we have a powerful
solver for the time dependent Schr\"odinger equation (TDSE) at our disposal \cite{scrinzi21:tRecX}. 
The time-dependent approach has the added benefit that one can explore the intensity
limits for perturbation theory and estimate the intensities that are admissible in a RABITT experiment.

Perturbation theory also simplifies calculations of the polarization dependence of the 
side-band: yields for all polarization directions of fundamental and harmonic
at given phase shift $\phi$ can be reconstructed from a maximum of 4 calculations at different polarizations
by the formula
\begin{align}
\Psi_{\al\be}(t)=&\cos\al\cos\be \Psi\up{zz}(t)+\sin\al\cos\be \Psi\up{xz}(t)
\nonumber\\
&+\cos\al\sin\be \Psi\up{zx}(t)+\sin\al\sin\be \Psi\up{xx}(t).
\label{eq:lincom}
\end{align}
where $\al$ and $\be$ are the polarization angles of fundamental and harmonic, respectively and polarization is assumed 
in the $xz$-plane. The $\Psi\up{ab}$ are 
\begin{equation}
\Psi\up{uv}(t)=\int_{-\infty}^t d\tau U_0(t-\tau)\sin{\om\tau}A\up{u}_f(t) U_0(\tau) \Phi\up{v}_{h}(\tau),\qquad u,v\in \{x,z\}.
\end{equation}
The subscripts refer to the respective polarization directions of fundamental and harmonic fields.
In the present paper we discuss only the case of parallel linear polarization $\vA_f\parallel\vA_h$, $\al=\be=:\ga$. 
In that case only the linear combination $\Psi\up{xz}+\Psi\up{zx}$ 
enters in Eq.~(\ref{eq:lincom}), such that wave packets $\Psi_{\ga\ga}$ computed at three different $\ga$
suffice for determining the result at all parallel polarization directions.

\subsection{The haCC model of the multi-electron system}

For the purpose of this study we consider the positions of nuclei as fixed 
and admit only single electronic excitation and single-electron emission. 
With these constraints one can make an ansatz for the $N$-electron
wave function $\Psi$ of the time-dependent Schr\"odinger equation at time $t$ as
\begin{equation}\label{eq:levelTop}
\Psi(\rangeSub{\vr}{1}{N};t)=\Psi_b(\rangeSub{\vr}{1}{N};t)+\sum_{c}\cA[\chi_c(\vr_1;t)\Phi_c(\rangeSub{\vr}{2}{N})], 
\end{equation}
where $\Psi_b$ is an in general time-dependent bound neutral wave function, $\Phi_c$ are stationary bound states
of the ion, which define the ionization channels, and $\chi_c$ are single-electron wave functions with bound 
as well as unbound content. For notational brevity, spin is subsumed into the $\vr_i$ coordinate. 
The notation $\cA[\ldots]$ indicates anti-symmetrization \wrt exchange of 
the electrons. Photo-electron spectra are obtained by asymptotically analyzing $\chi_c$ at large times   
in terms of scattering states. In practice one may freeze $\Psi_b$
at the neutral ground state, admitting only the single-electron excitations that are contained in $\cA[\chi_c\Phi_c]$
and limit the emission channels to a small number. This circumscribes the ``hybrid anti-symmetrized Coupled Channels'' (haCC)
approach of Ref.~\cite{majety15:hacc}. 

The haCC ansatz was used previously for computing ionization in a variety of non-perturbative 
strong field problems, where important effects of exchange and correlation were observed \cite{majety15:exchange,majety17:co2spectra}.
RABITT in contrast is inherently perturbative, but in turn the XUV emission is more sensitive
to the short range behavior of the solutions, such that conclusions from the strong-field regime cannot be carried over
to the present problem.

In the present paper we discuss ionization into the HOMO, HOMO-1, and HOMO-2 channels. These are for ionic states where the 
$X^2\Pi_g$, $A^2\Pi_u$ and $B^2\Si_u$ orbitals, respectively, are removed, further abbreviated as $X,A$ and $B$. We use real 
$\Pi$ orbitals, where orbitals with the node in the polarization plane are labeled as $X_n,A_n$ and the ones with an anti-node 
as $X_a,A_a$. The polarization plane is $xz$, 
i.e.\ at azimuthal angle $\phi=0$. 
In experiments, a statistical mixture of the two states would be observed, but we keep the
channels apart here for additional illustration of channel sensitivity.

\subsection{Simplified models}

While the {\em ab initio} approach above can provide the correct numbers, it does little to clarify the mechanisms at work.
Therefore we explore to which extent the approach may be simplified using more intuitive models. We use a hierarchy of models 
with Hartree-Fock as the most complete one,
a Hartree model discarding anti-symmetrization, and a set of single-electron models with the ``strong field approximation'' (SFA)
as its simplest representative,  whose only system-specific input is the electron's initial orbital. 

In all models, the electron is emitted from the same initial orbital. For that we 
separate the orbital $\varphi_c$ from the rest of the space by writing
\begin{equation}\label{eq:hamSubspace}
H(t)=|\varphi_c\r E_c \l \varphi_c| +Q H_c Q + i\vA(t)\cdot\vna,\quad Q:=1-|\varphi_c\r\l \varphi_c|
\end{equation}
The channel-specific Hamiltonian $H_c$ and the projector $Q$ are adjusted for the various models, while the initial energy
$E_c$ is kept at the known value for the given channel.

\subsubsection{A Hartree model for RABITT}

The Hartree ansatz is a reduction of the haCC ansatz (\ref{eq:levelTop}), 
where anti-symmetrization $\cA[\ldots]$ is omitted
and product states are used for $\Psi_b$ and $\Phi_c$. Within the perturbative conditions of RABITT, 
the harmonic field induces transitions to the continuum, and the fundamental modulates the continuum as to produce side bands. 
Under these conditions, any impact of channel coupling on the two-photon side band only appears in processes that involve more than two photons.
Consequently, we will consider all but a one orbital as frozen. 

We use for $\Psi_b$ a product of single electron orbitals $\varphi_c(\vr)$ 
and the ionic hole states $\Phi_c=a_c\Psi_b$, where $a_c$ denotes removal of $\varphi_c$. 
Integrating over the ionic degrees of freedom one obtains, in atomic units, single-electron Hamiltonians $H_c(t)=-\De/2+W_c$ for channel $c$,
with the nuclear plus Hartree potentials 
\begin{equation}\label{eq:potHartree}
W_c(\vr) =  V(\vr)+\sum_{c\neq c'}\int_\RR d^3r'\,\frac{|\varphi_{c'}|^2}{|\vr-\vr'|}.
\end{equation}
A plausible choice for the $\varphi_c$ are the energetically lowest Hartree-Fock molecular orbitals from the CI representations 
of $\Psi_b$ and $\Phi_c$ in (\ref{eq:levelTop}).
This Hartree ansatz is our basis for exploring the relevance of common approximations or improvements of the model
for attosecond delays. The interaction with the laser field $i\vA\cdot\vna$ acts on the complete space and in particular 
connects the initial orbital $\varphi_c$ to the rest of the single-electron space.

\subsubsection{A Hartree-Fock model for RABITT}
\label{sec:hartree_fock}
The HF ansatz extends the Hartree ansatz by full anti-symmetrization of the wave function.
It is related to Eq.~(\ref{eq:levelTop}) by using the HF ground state for $|\Psi\r$ and the channel functions
$|\Phi_c\r = a_c |\Psi_b\r $, where $a_c$ is the annihilation operator for $\varphi_c$. By symmetry,
none of the occupied orbitals $\varphi_c'$ is accessible for the ejected electron, which amounts to 
extending the projector $Q$ in Eq.~(\ref{eq:hamSubspace}) to all orbitals as
\begin{equation}\label{eq:constraintHF}
Q_{HF}=1-\sum_c|\varphi_c'\r\l \varphi_c'|.
\end{equation}
This reduces the dynamical freedom of the emitted electron comparing to the Hartree ansatz and creates an additional 
``nonlocal potential''.

The main technical complication of HF calculations in general is the appearance of an exchange operator,
which couples different ionization channels.
Within our assumption of all electrons in their Hartree-Fock orbitals except for laser-induced perturbations, 
this appears only in the next perturbative order beyond the two-photon limit
underlying RABITT. Therefore, on this lowest level of perturbation theory 
exchange can be neglected and we are left with the orthogonalization $Q_{HF}$
to all $\phi_c'$ as the only  effect of anti-symmetrization.

\subsubsection{SFA-like models with short and long-range potentials}

Models that further simplify the Hamiltonian~(\ref{eq:hamSubspace}) have been of great use for understanding the
physics of photoionization in general and of strong field physics in particular.
In SFA one assumes that the electron is in some initial bound state $\varphi_c$, but 
otherwise electronic motion in the laser field is free. With a few extra approximations, one can solve the system with analytic methods
(or very nearly so). Avoiding these additional approximations, we solve the TDSE where motion outside $\varphi_c$ is free,
i.e. we choose $H_c=-\De/2$. Analytic solutions usually disregard the need for orthogonalization by $Q$ 
when accounting for the dynamics after ionization, but still produce results very similar to the Hamiltonian (\ref{eq:hamSubspace}).

The key approximation of SFA --- the absence of any potential --- 
is also its main source of error. Clearly, the  potential introduces time-delays in the emission. 
Delays by the potential are commonly separated into 
delays by the short range part of the potential, that could be considered more genuine, and a universal
contribution due to the long-range Coulomb potential \cite{Dahlstrom12:delays}. One can analyze the relative importance of the
contributions using the the Hamiltonian (\ref{eq:hamSubspace}),
where one adds a potentials with different long- and short-range behavior, 
while keeping initial state $|\varphi_c\r$ and binding energy $E_c$ unchanged.

We are interested in effects of short-range interactions, of the Coulomb potential, and possibly
of longer range non-Coulombic parts of the potential. Somewhat arbitrarily we place the boundary 
between short- and long-range at a distance of $\approx 4$ atomic units from the $C$-atom in $CO_2$, which is a bit 
less than twice the $C\!-\!O$ bond length of $2.197\au$. We model these three cases by adding different potentials $QVQ$ outside
the initial state in (\ref{eq:hamSubspace}) that are purely Coulombic, molecular at short range and Coulombic at intermediate range,
or Coulombic at short range and molecular at intermediate range:
\begin{eqnarray}
V_q&=&-\frac{1}{r}\\
V_s&=&m(r) W_c(\vr)-\frac{1-m(r)}{r} \label{eq:Vshort}\\
V_\ell&=&[1-m(r)] W_c(\vr) - \frac{m(r)}{r} \label{eq:Vlong}.
\end{eqnarray}
Here $m(r)$ is a mask function that is $\equiv1$ for $r<R_m$ and $\equiv0$ for $r>R_n$ with a differentiably smooth 
transition from 1 to 0 on $[R_m,R_n]$.

\section{Numerical results and discussion}

All calculations presented below are based on the same set of orbitals $\varphi_c$ expanded in a single-center basis using spherical harmonics
\begin{equation}
\varphi_c(\vr)=\sum_{l=|m|}^L Y_l^{m}(\phi,\th) R_{lm}(r),
\end{equation}
where for $R_{ml}(r)$ a finite-element discrete variable (FE-DVR) representation 
is used with $m=0$ and $m=\pm1$ for the $\Si$ and $\Pi$ orbitals, respectively. 
The same expansion is used for the $\chi_c$ with a range of $m=-3,\ldots,3$, which accounts for two-photon transitions
from the initial $|m|\leq1$ states. Photo-electron spectra are computed using tSurff  \cite{Tao2012,scrinzi12:tsurff}. 
Computations were performed with the tRecX code as reported in Ref.~\cite{scrinzi21:tRecX}, where the reader can find details on computational implementation and an overview of computational methods.

The single-center spherical expansion is notoriously poorly convergent for a multi-center molecule, 
but the discretization errors do not affect conclusions of the relative importance
of short- and long-range effects and of the orthogonalization by Eq.~(\ref{eq:constraintHF}), if the 
same expansion is used across all models. We used a comparatively small number of $L=9$ for all model calculations,
but verified that the conclusions do not change with a larger value of $L=29$. 

For computing spectra we use tSurff with radius $R_c=35$, which implies that beyond that radius all effects of the 
molecular potential are neglected \cite{Tao2012,scrinzi12:tsurff}. The cutoff is well beyond the 
$C-O$ bond distance of $2.197\au$, such that we are in the Coulombic tail of 
the molecular potential and safely outside the reach of any occupied orbital $\varphi_c$. 
Note that absolute delays between Coulomb and free motion cannot be defined,
as these diverge logarithmically with the propagation distance. 
Rather, in our comparisons we assess the effects at short- and medium range interactions, where ``medium range'' means
distances where non-Coulombic contributions to the potential would be well-approximated by multipole potentials. 

The discretization in the radial coordinate is dense in the vicinity of the Coulomb singularities 
with a total of 40 functions on the interval $[0,5]$, which is followed by elements of size 5 up to $R_c$ with  FE-DVR order of $15$ 
on each. That radial discretization is fully converged, leaving the limitation in angular momenta as the main discretization error.
Outside $R_c$ we use irECS \cite{scrinzi10:irecs} for absorption, which implements exterior complex scaling using an exponentially 
damped polynomial basis. This gives perfect absorption in the present setting. 

\begin{figure}
\includegraphics[width=0.7\textwidth]{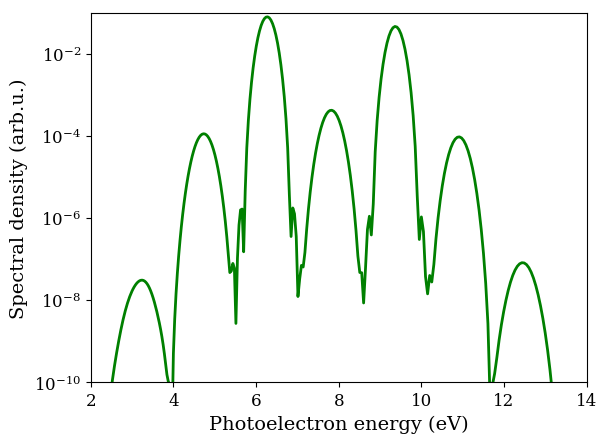}
\caption{\label{fig:spectrum} Photo-ionization peaks into the HOMO channel of $CO_2$ for an 800 nm pulse at $10^{10}\,W/cm^2$ combined with the 
13$^{th}$ and 15$^{th}$ harmonic at $10^{11}\,W/cm^2$. The side-band at $8\,eV$ between the two dominating harmonic peaks is used for RABITT. 
Calculation for the Hartree-Fock model.
 }
\end{figure}

The computation of photoelectron spectra by tSurff is numerically robust with rapid convergence and negligible numerical noise
such that equations (\ref{eq:twoPhoton}) and 
(\ref{eq:lincom}) are fulfilled faithfully for any choice of phases $\phi$ and polarization angles $\al,\be$. This agreement also shows that
for the given system the selected intensities of $10^{10}\,W/cm^2$ for the fundamental and $10^{11}\,W/cm^2$ for the harmonics
are safely in the perturbative range. The example spectrum 
in Fig.~\ref{fig:spectrum} shows harmonic peaks and side band peaks over a wide dynamic range.

\subsection{Hartree and Hartree-Fock}

\begin{figure}
\includegraphics[width=0.7\textwidth]{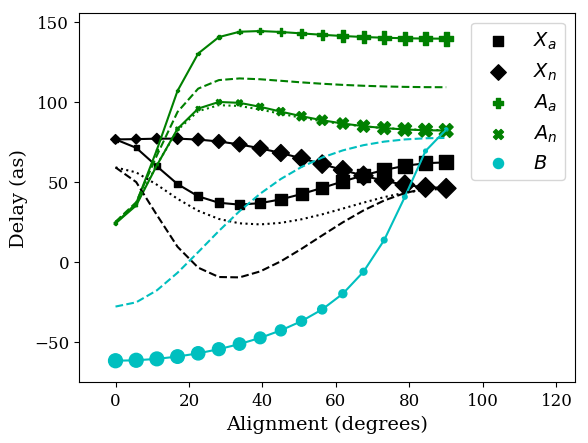}
\caption{\label{fig:hartree_fock}
Alignment-dependence of delays in Hartree-Fock description of $CO_2$ for the HOMO ($X^2\Pi_g$), HOMO-1 ($A^2\Pi_u$) and HOMO-2 ($C^2\Sigma_u$)
channels. The degenerate $\Pi$ states are chosen with a node ($X_n,A_n$) and anti-node  ($X_a,A_a$) in the polarization plane, respectively. 
Size of the plot symbols indicates the relative yields within each curve. The dashed and dotted lines matching the colors of the respective 
channels are obtained in Hartree-approximation, where dotted refers to $X_a$ and $A_a$, respectively.  
 }
\end{figure}

The Hartree model Eqs.~(\ref{eq:hamSubspace}) and (\ref{eq:potHartree}) is arguably the most accessible for an intuitive interpretation, 
as system-specific properties appear only in the initial orbital and as single-electron motion in a potential. We expect it 
to represent --- at least on a qualitative level --- the behavior of emission delays. Its minimally required extension is to 
Hartree-Fock on the perturbative level by replacing $Q$ with $Q_{HF}$, Eq.~(\ref{eq:constraintHF}).
Fig.~\ref{fig:hartree_fock} shows attosecond delays for ionization from the HOMO, HOMO-1 ($\Pi_g$ and $\Pi_u$)  
and HOMO-2 ($\Si_g$) orbitals. As to be expected, orientation dependence of yields is opposite for the $\Si$ and $\Pi$ symmetries, with 
the yield  in parallel alignment being maximal for the $\Si$-state and minimal for the $\Pi$ states. Delay times of the 
channels differ significantly, and also alignment dependence has different characteristics. It is important to observe 
that in Hartree approximation absolute delays change by up to 50 as or more 
relative to Hartree-Fock, which partly appears as an overall offset, but also 
as a modulation of alignment dependence. There appears to be a preponderance for anti-symmetrization to reduce alignment 
dependence, which might be interpreted as an effective screening of the Hartree potential by excluding the electron 
from the more attractive regions, but such an observation would need
a more systematic confirmation with a larger set of channels and more molecular species.
The important conclusion of this comparison is that any interpretation of delays on the
level of a few tens of attoseconds must include anti-symmetrization. 

\begin{figure}
\includegraphics[width=0.7\textwidth]{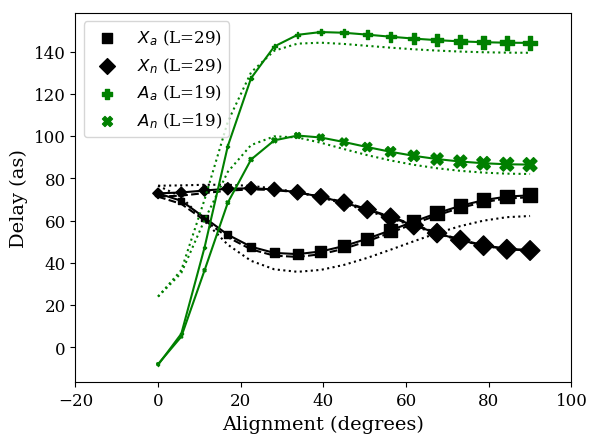}
\caption{\label{fig:convergeL} Convergence of delays in the HOMO and HOMO-1 channels for 10, 20 and 30 angular momenta. The solid
lines with markers are for maximal angular momenta, dotted lines for 10. Largest deviations appear at parallel alignment where yields are low.
For the HOMO, results with 20 angular momenta are also included, but not distinguishable 
from the converged curve on the resolution of the graph.}
\end{figure}

The figures shown were obtained using only 10 angular momenta in our single-center expansion of the wave function. While this is clearly
insufficient for fully converged results, it does not affect our conclusions. Figure \ref{fig:convergeL} shows the 
convergence with up to 30 angular momenta without qualitative change of the results.

\subsection{Short- and long-range scattering}

One expects that delays originate from scattering in the immediate vicinity of the nuclei roughly on the scale 
where the respective orbitals are located. We identify the relevant range of the potential by computing
delays for the screened Hartree potentials of Eqs.~(\ref{eq:Vshort}) and (\ref{eq:Vlong}). 
With the finding above that anti-symmetrization reduces the effect of the potential and alignment dependence, 
we include the results with and without the anti-symmetrization.  

\begin{figure}
\includegraphics[width=0.7\textwidth]{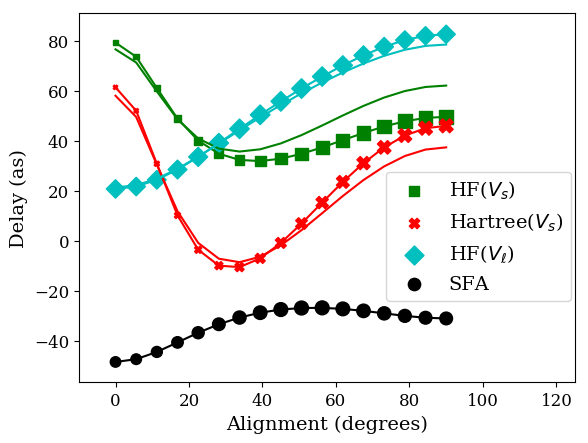}
\caption{\label{fig:models}
Spatial origin of RABITT delays in emission from the HOMO. Truncated $V_s$, Eq.~(\ref{eq:Vshort}), in
HF (green squares) and Hartree (red crosses) approximations. Lines next to the symbols are with the full Hartree potential.
Cyan diamonds: long range potential $V_\ell$, Eq.~(\ref{eq:Vlong}), Coulomb is the line next to it. 
Bullets are for zero potential without anti-symmetrization (SFA).}
\end{figure}

Fig.~\ref{fig:models} summarizes the results for these models. We see that
within accuracy $\sim 10\, as$, delays are caused by $V_s$, 
if we define the transition form short to long range in the interval [3,5].
Conversely, delays due to the long range potential $V_\ell$ are barely distinguishable from purely Coulombic delays.
This rapid transition of the potential to Coulombic behavior is to be expected for a non-polar molecule.

Even in absence of any potential in the continuum (SFA) one observes alignment-dependent delays. 
As the Fourier transform of our orbitals is strictly real, the phase shift is solely due to the projector 
$Q$ acting as a non-local potential, which is the minimal 
tribute one must pay in a consistent theory to the presence of an spatially extended initial bound state.

\subsection{Correlation and exchange}

Knowing that attosecond delays are highly sensitive to short-range modulations of the single electron potentials, an important
role of correlation is to be expected. The simplest manifestation of correlation is the modification of the initial state on 
a scale that one finds in Dyson orbitals. As short-range modulations are most important for XUV ionization and delays, admixtures
of tightly bound orbitals may impact delays. 

The most prominent effect of correlation comparing to the Hartree and Hartree-Fock description discussed so far is that
channels become coupled. In Ref.~\cite{kamalov20:CO2} channel coupling was identified as crucial for the description
attosecond delays. Physically this may be expected, as correlation means that a given electron is influenced by the specific
positions of other electrons, rather than just by their average field, again implying local modifications of forces
on every electron. 

\begin{figure}
\includegraphics[width=0.7\textwidth]{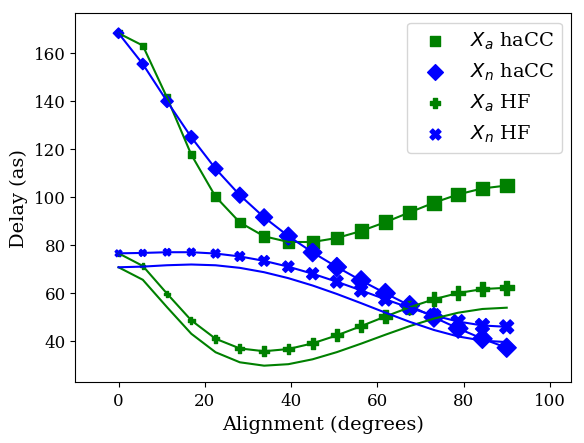}
\caption{\label{fig:hacc} The importance of correlation in photoemission delays for the X channels. Squares: haCC, Eq.~(\ref{eq:levelTop}), 
bullets: Hartree-Fock, lines without symbols: Hartree-Fock using the Dyson orbital for the initial orbital.
}
\end{figure}

The haCC ansatz (\ref{eq:levelTop}) admits electron correlation on a CI level and fully includes exchange.
For the present calculations we describe the neutral initial state $\Psi_b$ and the lowest 6 ionic states $\Phi_c$ 
on the CI singles doubles level with an atom centered minimally augmented cc-pvtz basis as obtained 
from the COLUMBUS package \cite{lischka11:columbus}. For further details on computation and method we refer the 
reader to Refs.~\cite{majety15:hacc,majety17:co2spectra}.
Fig.~\ref{fig:hacc} shows delays in the  two degenerate HOMO channels as obtained with and without exchange 
and correlation. The degenerated channels are distinguished by the dominant channel orbitals
with a node ($X_n$) or anti-node ($X_a$) in the polarization plane, respectively. We observe some modification of yields
and dramatic changes in the computed delays. In all cases yields are minimal at parallel alignment and
increase monotonically towards perpendicular alignment, the increase is less pronounced for the haCC calculation.
Delays in the $X_a$ channel are minimal around 40 degrees alignment while delays
in the $X_n$ channel decrease monotonically with the aliment angle. 

On the scale of the observed effects, initial state correlation as represented by a Dyson orbital is negligible:
Fig.~\ref{fig:hacc} includes a HF calculation with $\varphi_c$ replaced with the corresponding
Dyson orbital. The small impact is consistent with the fact, that the most important admixtures in the Dyson orbital
is from more diffuse orbitals, while all contributions from more tightly bound orbitals remain below the level
of $10\inv5$.

It must be mentioned that due to the full inclusion of exchange, haCC computations are very resource intensive and
systematic convergence studies have not been possible yet. 
While pushing convergence of the haCC calculation may change results on the level of many tens of attoseconds, any
convergence towards closer resemblance with the Hartree-Fock calculation would be purely accidental: as discussed 
in Sec.~\ref{sec:hartree_fock}, Hartree-Fock appears as the limiting case of haCC in the independent particle picture. 
A systematic convergence study of the haCC calculation to assess the reliability of
haCC delays {\it ab initio} will be presented separately.

\section{Summary and conclusions}

Phase shifts in RABITT setups that correspond to delays of 10 or more attoseconds were found to depend on 
every detail of modeling of the physical process. This includes elementary aspects like the exact shape of 
the effective potential seen by the emitted electron, which precludes any approximation that does not fully
take into account electron screening on the level of the Hartree potential. Similarly, the simple 
constraint of the emitted electron's dynamics by Pauli exclusion must be imposed. 
The far dominant effects appear locally in the vicinity of the nuclei, which explains the pronounced 
effect of correlation on delays seen in the present simulation. Correlation effects are also highlighted 
by the importance of channel coupling reported earlier in Ref.~\cite{kamalov20:CO2}.
In view of this high sensitivity, one must also examine the validity the common frozen nuclei approximation when 
computing of delays with high accuracy.  

It appears, that any theory-experiment comparison of delays on the level of a few tens of attoseconds requires 
{\it ab initio} level theory. This casts doubt on the usefulness of attosecond delay measurements for 
independent insight into electronic or nuclear dynamics in molecules. On the other hand, it establishes an outstanding 
scientific challenge and a highly sensitive test for fully correlated multi-electron calculations 
and the matching precision experiments.

\bibliographystyle{unsrt}
\bibliography{/home/scrinzi/Papers/bibliography/photonics_theory}

%\printbibliography

\end{document}

%% file: my_commands.tex
\usepackage{amsmath}
\usepackage{amssymb}
\usepackage{makeidx}
\usepackage{graphicx}
\usepackage{color}

\newcommand{\wrt}{w.r.t.\ }

\newcommand{\hide}[1]{}

\newcommand{\up}[1]{ ^{(#1)}}
\newcommand{\inv}[1]{ ^{-#1}}
\newcommand{\myvec}[1]{\vec{#1}}
\newcommand{\vr}{{\myvec{r}}}

\newcommand{\vA}{{\myvec{A}}}

\newcommand{\rangeSub}[3]{#1_#2,\ldots,#1_{#3}}

\newcommand{\cA}{\mathcal{A}} % anti-symmetrization
\newcommand{\RR}{\mathbb{R}}%{ {I\!\!R} }
\renewcommand{\r}{\rangle}
\renewcommand{\l}{\langle}

\newcommand{\om}{\omega}

\newcommand{\Si}{\Sigma}

\newcommand{\al}{\alpha}
\newcommand{\la}{\lambda}

\newcommand{\be}{\beta}

\newcommand{\ga}{\gamma}

\newcommand{\De}{\Delta}
\renewcommand{\th}{\theta}

\newcommand{\lcase}{\left\{\begin{array}{ll}}
\newcommand{\rcase}{\end{array}\right.}
\renewcommand{\bar}{\begin{array}{ll}}
\newcommand{\ear}{\end{array}}
\newcommand{\bal}{\begin{align}}
\newcommand{\eal}{\end{align}}
\newcommand{\bma}{\begin{pmatrix}}
\newcommand{\ema}{\end{pmatrix}}
\newcommand{\beq}{\begin{equation}}
\newcommand{\eeq}{\end{equation}}
\newcommand{\bel}[1]{\begin{equation}\label{eq:#1}}
\newcommand{\eel}{\end{equation}}
\newcommand{\bea}{\begin{eqnarray}}
\newcommand{\eea}{\end{eqnarray}}
\newcommand{\beaNN}{\begin{eqnarray*}}
\newcommand{\eeaNN}{\end{eqnarray*}}

\newcommand{\vEf}{\vec{\mathcal{E}}}

\newcommand{\vna}{\vec{\nabla}}

%% file: VolkmannCO2.bbl
\begin{thebibliography}{10}

\bibitem{Yakovlev2005}
Vladislav Yakovlev, Ferdinand Bammer, and Armin Scrinzi.
\newblock {Attosecond streaking measurements}.
\newblock {\em Journal of Modern Optics}, 52(2):395--410, January 2005.

\bibitem{veniard96:phase-dependence}
Val\'erie V\'eniard, Richard Ta{\"i}eb, and Alfred Maquet.
\newblock Phase dependence of (n+1)-color (n$>$1) ir-uv photoionization of
  atoms with higher harmonics.
\newblock {\em Phys. Rev. A}, 54:721--728, 1996.

\bibitem{paul01:atto}
P.~M. Paul, E.~S. Toma, P.~Breger, G.~Mullot, F.~Auge, Ph. Balcou, H.~G.
  Muller, and P.~Agostini.
\newblock Observation of a train of attosecond pulses from high harmonic
  generation.
\newblock {\em Science}, 292:1689--1692, 2001.

\bibitem{P.Eckle2008}
{P. Eckle}, A~N Pfeiffer, C~Cirelli, A~Staudte, R~Doerner, H~G Muller,
  M~B\"{u}ttiker, and U~Keller.
\newblock {Attosecond Ionization and Tunneling Delay Time Measurements in
  Helium}.
\newblock {\em Science}, 322(December):1525--1529, 2008.

\bibitem{sainadh19:attodelays}
U.~Satya Sainadh, Han Xu, Xiaoshan Wang, A.~Atia-Tul-Noor, William~C. Wallace,
  Nicolas Douguet, Alexander Bray, Igor Ivanov, Klaus Bartschat, Anatoli
  Kheifets, R.~T. Sang, and I.~V. Litvinyuk.
\newblock {Attosecond angular streaking and tunnelling time in atomic
  hydrogen}.
\newblock {\em {Nature}}, {568}({7750}):{75+}, {Apr 4} {2019}.

\bibitem{BuettikerM.andLandauer1982}
R.~{Buettiker, M. and Landauer}.
\newblock {Traversal time for tunneling}.
\newblock {\em Physical Review Letters}, 49(23):1739--1742, 1982.

\bibitem{torlina15:attoclock}
L.~Torlina, F.~Morales, J.~Kaushal, I.~Ivanov, A.~Kheifets, A.~Zielinski,
  A.~Scrinzi, H.G. Muller, S.~Sukiasyan, M.~Ivanov, and O.~Smirnova.
\newblock Interpreting attoclock measurements of tunnelling times.
\newblock {\em Nature Physics}, 11:503--508, 2015.

\bibitem{Argenti17:delays}
L.~Argenti, \'A. Jim\'enez-Gal\'an, J.~Caillat, R.~Ta\"{\i}eb, A.~Maquet, and
  F.~Mart\'{\i}n.
\newblock Control of photoemission delay in resonant two-photon transitions.
\newblock {\em Phys. Rev. A}, 95:043426, Apr 2017.

\bibitem{saalmann20:delays}
Ulf Saalmann and Jan~M. Rost.
\newblock Proper time delays measured by optical streaking.
\newblock {\em Phys. Rev. Lett.}, 125:113202, Sep 2020.

\bibitem{kamalov20:CO2}
Andrei Kamalov, Anna~L. Wang, Philip~H. Bucksbaum, Daniel~J. Haxton, and
  James~P. Cryan.
\newblock Electron correlation effects in attosecond photoionization of
  {$CO_2$}.
\newblock {\em Phys. Rev. A}, 102:023118, Aug 2020.

\bibitem{Pazourek15:atto}
Renate Pazourek, Stefan Nagele, and Joachim Burgdoerfer.
\newblock {Attosecond chronoscopy of photoemission}.
\newblock {\em Rev. Mod. Phys.}, {87}({3}), Aug 12 {2015}.

\bibitem{scrinzi21:tRecX}
A.~Scrinzi.
\newblock {tRecX} --- an environment for solving time-dependent
  {S}chr\"odinger-like problems.
\newblock (preprint) {https://arxiv.org/abs/2101.08171}.

\bibitem{majety15:hacc}
Vinay~Pramod Majety, Alejandro Zielinski, and Armin Scrinzi.
\newblock {Photoionization of few electron systems: a hybrid coupled channels
  approach}.
\newblock {\em {New. J. Phys.}}, {17}, {Jun 1} {2015}.

\bibitem{majety15:exchange}
Vinay~Pramod Majety and Armin Scrinzi.
\newblock Dynamic exchange in the strong field ionization of molecules.
\newblock {\em Phys. Rev. Lett.}, 115:103002, Sep 2015.

\bibitem{majety17:co2spectra}
Vinay~Pramod Majety and Armin Scrinzi.
\newblock Multielectron effects in strong-field ionization of {$CO_2$}: Impact
  on differential photoelectron spectra.
\newblock {\em Phys. Rev. A}, 96:053421, Nov 2017.

\bibitem{Dahlstrom12:delays}
J~M Dahlstr\"{o}m, A~L'Huillier, and A~Maquet.
\newblock Introduction to attosecond delays in photoionization.
\newblock {\em Journal of Physics B: Atomic, Molecular and Optical Physics},
  45(18):183001, Aug 2012.

\bibitem{Tao2012}
Liang Tao and Armin Scrinzi.
\newblock {Photo-electron momentum spectra from minimal volumes: the
  time-dependent surface flux method}.
\newblock {\em New Journal of Physics}, 14(1):013021, Jan 2012.

\bibitem{scrinzi12:tsurff}
Armin Scrinzi.
\newblock t-{\protect surff}: fully differential two-electron photo-emission
  spectra.
\newblock {\em New Journal of Physics}, 14(8):085008, 2012.

\bibitem{scrinzi10:irecs}
Armin Scrinzi.
\newblock Infinite-range exterior complex scaling as a perfect absorber in
  time-dependent problems.
\newblock {\em Phys. Rev. A}, 81(5):053845, May 2010.

\bibitem{lischka11:columbus}
Hans Lischka, Thomas Mueller, Peter~G. Szalay, Isaiah Shavitt, Russell~M.
  Pitzer, and Ron Shepard.
\newblock {COLUMBUS --- a program system for advanced multireference theory
  calculations}.
\newblock {\em {Wiley Interdiscip. Rev.-Comp. Mol. Science}}, {1}({2}):{191},
  {2011}.

\end{thebibliography}
